\documentclass[preprint,3p,12pt]{elsarticle}
\usepackage{mathrsfs}
\usepackage{amsmath}
\usepackage{stmaryrd}
\usepackage{bbding}
\usepackage{dcolumn}
\usepackage{graphicx}	
\usepackage{amsfonts}
\usepackage{amssymb}
\usepackage{psfrag}
\usepackage{wrapfig}
\usepackage{subfigure}
\usepackage{makeidx}
\usepackage{bm}
\usepackage{epsf}
\usepackage{epsfig}
\usepackage{setspace}
\usepackage{graphicx}
\usepackage{epstopdf}
\usepackage{psfrag}
\usepackage{subfigure}
\usepackage{color}
\usepackage{comment}
\usepackage{caption}
\renewcommand{\vec}[1]{{\boldsymbol{#1}}}

\begin{document}

\title{Modeling Equations in Wave-Particle Turbulence Simulation}

\author[HKUST1]{Xiaojian Yang}
\ead{xyangbm@connect.ust.hk}

\author[HKUST1]{Gaocheng Liu}
\ead{gliuau@connect.ust.hk}

\author[HKUST1,HKUST2,HKUST3]{Kun Xu\corref{cor1}}
\ead{makxu@ust.hk}

\address[HKUST1]{Department of Mathematics, Hong Kong University of Science and Technology, Clear Water Bay, Kowloon, Hong Kong}
\address[HKUST2]{Department of Mechanical and Aerospace Engineering, Hong Kong University of Science and Technology, Clear Water Bay, Kowloon, Hong Kong}
\address[HKUST3]{Shenzhen Research Institute, Hong Kong University of Science and Technology, Shenzhen, China}
\cortext[cor1]{Corresponding author}

\begin{abstract}

Recently, the wave-particle turbulence simulation (WPTS) has been proposed as a novel framework for non-equilibrium turbulence modeling and simulation. In this work, for the first time the complete model equations of WPTS are explicitly derived from the perspective of wave-particle decomposition, and the physical mechanism of each term is clearly interpreted.
To extend its applicability to wall-bounded flows, the WPTS coupled with wall model is developed, and the introduction of wall model substantially alleviates the near-wall grid-resolution constraint. In the bulk region, the wave component resolves the large-scale structures, whereas the particle component accounts for subgrid-scale modeling through the non-equilibrium transport mechanism. As a result, the coupled method enables accurate predictions of the flat-plate transition on coarse-grid. In particular, the computed skin-friction coefficient and mean velocity profiles in the fully turbulent region agree well with the reference data from direct numerical simulation, and the accuracy is markedly superior to that of the gas-kinetic scheme (GKS) under the identical grid. These findings underscore the considerable promise of the multi-scale WPTS method for transitional flow simulations.

\end{abstract}


\maketitle

\section{Introduction}

The distinguished feature of turbulent flow is the presence of interconnected flow structures spanning an extensive range of scales, which brings fundamental challenges to the development of numerical methods. By solving the Navier–Stokes (NS) equations, the conventional methods include the direct numerical simulation (DNS), the large eddy simulation (LES), and the Reynolds-averaged Navier–Stokes (RANS) approach \cite{nieuwstadt2016turbulence, leschziner2015statistical}.
In recent years, growing interests focus on the kinetic theory for turbulence modeling and numerical method development \cite{Tur-kinetic-xin2026model, Tur-kinetic-luan2025constructing}. Notably, the wave-particle turbulence simulation (WPTS) offers a novel pathway \cite{Tur-wpts-first-yang2025wave, Tur-wpts-second-yang2025wave}. 
In WPTS, the distribution function characterizing the fluid state is decomposed into wave and particle components, and these two components are coupled evolved under the kinetic framework. The construction of WPTS originates from the the unified gas-kinetic scheme (UGKS) and particularly its wave-particle version, unified gas-kinetic wave-particle (UGKWP) method, which are developed for the problems with non-equilibrium transport mechanism \cite{UGKS-xu2010unified, WP-first-liu2020unified}. The objective of WPTS is the multi-scale modeling for turbulent flow spanning from the NS-resolved scales up to sub-grid scales that cannot be fully resolved on coarse meshes. 

One key advantage of WPTS lies in the fact that the proportion of particle component adjusts automatically in response to the local turbulent intensity. Specifically in laminar flow regimes, no particles are generated, and WPTS automatically reduces to the gas-kinetic scheme (GKS), a kinetic NS solver \cite{GKS-2001}. As the turbulent intensity increases to the situation that the employed grid becomes locally insufficient, the weight of particles increases. As a result, WPTS constitutes a unified framework capable of adaptively describing both laminar and turbulent flow, indicating it naturally suitable for problems involving multiple flow regimes.
The flat-plate transition problem is one representative case, where the flow originates from a laminar state, undergoes disturbance growth, and eventually transitions to a full turbulent state \cite{Tur-case-plate-pirozzoli2004direct, Tur-case-plate-hui2005direct, Tur-case-plate-zhou2019subgrid}. Therefore, it instantaneously encompasses laminar, transitional, and fully turbulent flow regimes at different locations, making it ideal for evaluating the multi-scale WPTS method. It constitutes the primary motivation for the present investigation.

This paper is organized as follows. In Section 2, the governing equations underlying WPTS is constructed and explained. Section 3 provides a detailed description of WPTS, along with its coupling with wall model. In Section 4, the performance of the wall-model WPTS is assessed through the flat-plate transition problem, and the last one is conclusion.

\section{Model equations of wave-particle turbulence simulation}

In this section, the model equation of WPTS is presented.
The typical feature in WPTS is wave-particle decomposition, and thus the total distribution function $f(\boldsymbol{x}, \boldsymbol{u}, t)$ of fluid elements is decomposed into wave and particle components, namely $f_w(\boldsymbol{x}, \boldsymbol{u}, t)$ and $f_p(\boldsymbol{x}, \boldsymbol{u}, t)$
\begin{equation}
	f(\boldsymbol{x}, \boldsymbol{u}, t) = f_w(\boldsymbol{x}, \boldsymbol{u}, t) + f_p(\boldsymbol{x}, \boldsymbol{u}, t).
	\label{eq:fdecomp}
\end{equation}
The decomposition in Eq.\eqref{eq:fdecomp} indicates the fluid elements are regarded as two types which will be evolved based wave and stochastic particles in WPTS. Particularly for $f_p(\boldsymbol{x}, \boldsymbol{u}, t)$, it can be obtained from discrete particles by the projection
\begin{equation}
	f_p(\boldsymbol{x}, \boldsymbol{u}, t) = 
	\frac{1}{V_\Omega} 
	\sum_{p \in \Omega} 
	\delta(\boldsymbol{x} - \boldsymbol{x}_p) \,
	\delta(\boldsymbol{u} - \boldsymbol{u}_p) \,
	m_p.
	\label{eq:fpgrid}
\end{equation}
The motivation of such a decomposition is multi-scale modeling of the turbulent flow under un-resolved gird. 
Specifically, $f_w$ represents the flow in the resolved scales, and $f_p$ represents the flow of unresolved scales. 
As a result, this decomposition depends on both the flow structure and the grid employed to represent the solution.

Then the kinetic model equation under the wave-particle decomposition framework will be given. 
Firstly, with the consideration of that the wave component targets the evolution of cell-resolved fluid motion, the kinetic model equation in WPTS is the following  
\begin{equation}
	\begin{aligned}
		\frac{\partial f_w}{\partial t} + \boldsymbol{u} \cdot \nabla_x  f_w 
		+ \boldsymbol{a}^{w} \cdot \nabla_{\boldsymbol{u}} f_w
		&= \frac{g - f}{\tau}
		+ \frac{f_p}{\tau + \tau_t}
		- \mathcal{S}_{gen}\left(f_w, f_p, \tau_t, h\right),
	\end{aligned}
	\label{eq:wave}
\end{equation}
where $f = f_w + f_p$, and $\left(g - f\right)/\tau$ is the relaxation term with the physical collision time $\tau$, i.e., $\tau = \mu/p$. 
The equilibrium distribution $g$ is the standard Maxwellian distribution
\begin{equation}
	g = \rho \left( \frac{\lambda}{\pi} \right)^{\frac{K+3}{2}} 
	\exp\left[-\lambda \left( (\boldsymbol{u} - \boldsymbol{U})^2 + \boldsymbol{\xi}^2 \right)\right],
	\label{eq:gw}
\end{equation}
where $\lambda = 1/\left(2R T\right)$, and $T$ is the total temperature \cite{Tur-wpts-first-yang2025wave}. The equilibrium state is assumed as the state where all the turbulent kinetic energy (TKE) (if existing) has been transferred to the thermal one by physical viscosity. 
The last two terms in Eq.~\eqref{eq:wave} are source terms for $f_w$, which are caused by the transformation between wave and particle component, namely the annihilation and generation of particles in WPTS. They also appear in the kinetic equation for the particle component in Eq.\eqref{eq:particle}, and will be introduced later.
On the left-hand side of Eq.~\eqref{eq:wave}, the term $\boldsymbol{a}^{w} \cdot \nabla_{\boldsymbol{u}} f_w$ represents the external forcing acting on the wave component, serving as the counterpart to the $\boldsymbol{a}^{p} \cdot \nabla_{\boldsymbol{u}} f_p$ for the particle component.

The non-equilibrium feature of WPTS is achieved mainly by the evolution of particle component, which differs WPTS from the viscosity-type turbulence modeling. The kinetic equation for particle component is modeled as
\begin{equation}
	\begin{aligned}
		\frac{\partial f_p}{\partial t} + \boldsymbol{u} \cdot \nabla_x f_p + \boldsymbol{a}^{p} \cdot \nabla_{\boldsymbol{u}} f_p
		&= \frac{-f_p}{\tau + \tau_{t}} + \mathcal{S}_{gen}\left(f_w, f_p, \tau_t, h\right)
	\end{aligned}
	\label{eq:particle}
\end{equation}
At the left hand side of Eq.\eqref{eq:particle}, $\boldsymbol{a}^{p} \cdot \nabla_{\boldsymbol{u}} f_p$ stands for the forcing term on the stochastic particles. In this paper, the pressure gradient force in particles' migration and the inter-particle interaction are taken into consideration, which will be introduced in details in next section.
The term $-f_p/\left(\tau + \tau_t\right)$ on the right-hand side of Eq.~(\ref{eq:particle}) stands for the decay of the particle component, which is associated with particle annihilation in the WPTS. In addition, $\mathcal{S}_{gen}\left(f_w, f_p, \tau_t, h\right)$ represents the increment of $f_p$ due to the particle sampling. Particularly, $\mathcal{S}_{gen} \left(f_w, f_p, \tau_t, h\right)$ depends on the modeled turbulence characteristic time $\tau_t$ and the numerical scale $h$ in resolving the local turbulent flow. For example, the proportion of newly-sampled particle is determined by $\tau_t$.  

In WPTS, the $f_w$ will not be updated explicitly, but $f$ is updated by multi-scale numerical flux constructed based on the wave-particle evolution, and further $f_p$ is updated directly through the evolution of stochastic particles.
Now we combine the model equations for wave component Eq.\eqref{eq:wave} and particle component Eq.\eqref{eq:particle}, and obtain
\begin{equation}
	\begin{aligned}
		\frac{\partial f}{\partial t} + \boldsymbol{u} \cdot \nabla_x f
		&= \frac{g- f}{\tau}.
	\end{aligned}
	\label{eq:kinetictot}
\end{equation} 

It should be emphasized that, in the equation governing $f$ Eq.\eqref{eq:kinetictot}, the term involving the external force on the particles and the corresponding counterpart term in the wave component are mutually canceled, leaving no explicit acceleration contribution. 
Importantly, the compatibility condition for the relaxation term in Eq.\eqref{eq:kinetictot} are satisfied
\begin{gather*}
	\int \vec{\psi}\left(g - f\right)\text{d}\vec{\Xi} = \vec{0}, ~~~
	\vec{\psi}=\left(1,\vec{u},\displaystyle \frac{1}{2}\left(\vec{u}^2+\vec{\xi}^2\right)\right)^T,
\end{gather*}
by which the conservation of mass, momentum and total energy can be preserved.

\section{Wave-Particle Turbulence Simulation (WPTS) Method}

For the kinetic equation Eq.\eqref{eq:kinetictot}, the integral solution can be obtained
\begin{equation}\label{bgk-integrasol}
	f(\vec{x},t,\vec{u})=\frac{1}{\tau}\int_0^t g(\vec{x}',t',\vec{u} )e^{-(t-t')/\tau}\text{d}t'\\
	+e^{-t/\tau}f_0(\vec{x}-\vec{u}t, \vec{u}),
\end{equation}
where $\vec{x}'=\vec{x}+\vec{u}(t'-t)$ is the trajectory of particles, $f_0$ is the initial gas distribution function at time $t=0$. 
Following the construction of GKS, we can obtain the time-dependent distribution function $f$ at the cell interface
\begin{align}\label{eqffullgks}
	f(\vec{x},t,\vec{u})
	=& c_1 g_0\left(\vec{x},\vec{u}\right)+ c_2 \overline{\vec{a}} \cdot \vec{u} g_0\left(\vec{x},\vec{u}\right)
	+c_3{\bar{A}} g_0\left(\vec{x},\vec{u}\right)\nonumber\\
	+& [c_4 + c_5 \vec{a}^r \cdot \vec{u} + c_6 A^r]g^{r} \left(\vec{x},\vec{u}\right)(1-H(u)) \nonumber\\
	+& [c_4 + c_5 \vec{a}^l \cdot \vec{u} + c_6 A^l]g^{l}\left(\vec{x},\vec{u}\right)H(u),
\end{align}
with the coefficient
\begin{align}\label{coeforgks}
	c_1 = 1-e^{-t/\tau}, ~~
	c_2 = \left(t+\tau\right)e^{-t/\tau}-\tau&, ~~
	c_3 = t-\tau+\tau e^{-t/\tau}, \nonumber\\
	c_4 = e^{-t/\tau}, ~~
	c_5 = -\left(t+\tau\right)e^{-t/\tau}&, ~~
	c_6 = -\tau e^{-t/\tau}.
\end{align}

The integral solution in Eq.\eqref{bgk-integrasol} illustrates that the evolution process comprises two aspects: the cumulative effect of the equilibrium state and the free transport of the initial distribution function, which is also the key of multi-scale numerical method.

For laminar flow the coefficients in Eq.\eqref{coeforgks} lead to the GKS, a second-order kinetic solver for NS solution, which is the limiting case of WPTS.
For turbulent flow, with the increasing of particle's proportion, the fluid elements in WPTS are divided into wave and particle two components. Importantly, the particle component plays key role in capturing non-equilibrium transport in turbulence. As a result, these Lagrangian particles are employed to model unresolved flow structures in turbulence on coarse computational grids \cite{Tur-wpts-first-yang2025wave}. For particles, the governing equation adopts a $\tau_t$-dependent relaxation formulation with the external force, namely
\begin{gather}\label{dudtpar}
	\frac{\text{d} \vec{u}\left(t\right)}{\text{d} t} = \frac{\vec{U} - \vec{u}}{\tau_n} + \vec{a}.
\end{gather}
In this paper, $\vec{a}$ is taken as $-\nabla p/\rho$, considering it as the dominant term.
The characteristic time $\tau_n$ in Eq.\eqref{dudtpar} is taken as the sum of physical and turbulent values, namely, $\tau_n = \tau + \tau_t$, where $\tau_t$ will be modeled based on the mixing-length hypothesis for the plate transition problem and introduced later.
Leveraging the UGKWP method, the collisionless particles (fluid parcels), which can survive a whole time step, are sampled and evolved via Lagrangian particles, and the flux contribution of remaining particles whose transport time are smaller than one time step will be calculated by the wave formulation, denoted as $\vec{F}^{fr,wave}$ \cite{WP-first-liu2020unified, Tur-wpts-first-yang2025wave}. Besides, in this paper the free transport flux, which is evaluated from the macroscopic variables and associated with the particles surviving at the end of previous step, is subtracted, leading to
\begin{align*}\label{eqFluxfrwave}
	\vec{F}^{fr,wave}_{ij}
	&=\vec{F}^{fr,UGKS}_{ij}\left(\vec{W}_i^{h}\right) 
	- \vec{F}^{fr,DVM}_{ij}\left(\vec{W}_i^{hp}\right)
	- \vec{F}^{fr,DVM}_{ij}\left(\vec{W}_i^{p,sur}\right)\\
	&=\frac{1}{\Delta t}\int_{0}^{\Delta t} \int \vec{u} \cdot \vec{n}_{ij} \left[ e^{-t/\tau}f_0(\vec{x}-\vec{u}t,\vec{u})\right] \vec{\psi} \text{d}\vec{u}\text{d}t\\
	&-e^{-\Delta t/\tau_n} \frac{1}{\Delta t} \int_{0}^{\Delta t} \int \vec{u} \cdot \vec{n}_{ij} \left[g_0^h\left(\vec{x},\vec{u} \right) - t\vec{u} \cdot g_{\vec{x}}^h\left(\vec{x},\vec{u} \right) \right] \vec{\psi}\text{d}\vec{u}\text{d}t\\
	&-\alpha \frac{1}{\Delta t} \int_{0}^{\Delta t} \int \vec{u} \cdot \vec{n}_{ij} \left[g_0^h\left(\vec{x},\vec{u} \right) \right] \vec{\psi}\text{d}\vec{u}\text{d}t\\
	&=\frac{1}{\Delta t} \int \vec{u} \cdot \vec{n}_{ij} \left[ \left(q_4 + q_4^{'} \right) g_0^h \left(\vec{x},\vec{u} \right)
	+ \left(q_5 + q_5^{'}\right) \vec{u} \cdot g_{\vec{x}}^h\left(\vec{x},\vec{u} \right) \right]\vec{\psi}\text{d}\vec{u},
\end{align*}
with
\begin{gather*}
	q_4=\tau\left(1-e^{-\Delta t/\tau}\right), ~~~
	q_4^{'} = - \Delta t e^{-\Delta t/\tau_n} - \alpha \Delta t, \\
	q_5=\tau\Delta te^{-\Delta t/\tau} - \tau^2\left(1-e^{-\Delta t/\tau}\right), ~~~
	q_5^{'} = \frac{\Delta t^2}{2}e^{-\Delta t/\tau_n},
\end{gather*}
where $\alpha$ is evaluated as $\alpha = \rho^{p,sur} / \rho^h$, and $\rho^{p,sur}$ is the density by counting the surviving particles at the end of last step. 

WPTS is based on the finite volume framework. Generally the evolution of wave and particle component will be summarized as below. 
On the one hand, the evolution for wave component mainly includes three steps:
\begin{itemize}
	
	\item {\bf Reconstruction}: 
	interpolating cell-averaged conserved variables $\vec{W}$ to interfaces. In this paper, the fifth-order WENO-AO reconstruction is employed.
	
	\item {\bf Evolution}: 
	computing fluxes $\vec{F}^{eq}$ by taking moment of $f^{eq}$
	\begin{align}\label{eqFluxeq}
		\vec{F}^{eq}_{ij}
		&=\frac{1}{\Delta t} \int_{0}^{\Delta t} \int \vec{u}\cdot\vec{n}_{ij} f_{ij}^{eq}(\vec{x},t,\vec{u})\vec{\psi}\text{d}\vec{u}\text{d}t \nonumber \\
		&=\frac{1}{\Delta t} \int_{0}^{\Delta t} \int \vec{u}\cdot\vec{n}_{ij} \left[ c_1 g_0\left(\vec{x},\vec{u}\right)
		+ c_2 \overline{\vec{a}} \cdot \vec{u} g_0\left(\vec{x},\vec{u}\right)
		+ c_3 A g_0\left(\vec{x},\vec{u}\right) \right] \vec{\psi}\text{d}\vec{u}\text{d}t.
	\end{align}
	
	\item {\bf Projection}: 
	updating the cell-averaged conserved variables $\vec{W}$ by $\vec{F}^{eq}$. Since both the equilibrium and free-transport fluxes are required in the projection, the complete formula for updating $\vec{W}$ will be given later in the detailed presentation of all flux contributions.
\end{itemize}

On the other hand, the evolution of stochastic particles is dominant in capturing the non-equilibrium transport in WPTS, and generally it can be summarized as: 
\begin{itemize}
	
	\item {\bf Generation}: 
	sample particles from the wave
	\begin{gather}\label{whp}
		\vec{W}^{hp}_i = e^{-\Delta t/\tau_n} \vec{W}^{h}_i,
	\end{gather}
	with $t_f = \Delta t$, and the velocity of the sampled particle is determined by
	\begin{gather}\label{velpart}
		\vec{u}_p = \delta \vec{u}_p + \vec{U},
	\end{gather}
	where $\delta\vec{u}_p = \mathcal{D}_{N} \left[\rho E_t^{prod}, \rho^h\right]$, depending on the fluid density by wave component $\rho^h$, and the modeled production of TKE. In this paper, it is assumed as $\rho E_t^{prod} = C_0 e^{-\Delta t/\tau_n} \rho e$ with $\rho e = \frac{1}{2}\rho \vec{U}^2$, and $C_0=1.0$ is employed. 
	More details about the particle sampling can be found in \cite{Tur-wpts-first-yang2025wave}.
	
	The spatial positions of particles within each cell are determined as follows. Initially, the initial position $\vec{x}$ of each particle is generated via uniform random sampling inside the cell. Subsequently, for each spatial dimension $i=1,2,3$, the coordinate component $x_i$ is either reflected symmetrically about the cell center $x_{c,i}$ or kept as the original value,
	which is based on the criterion that the resulting particle spatial distribution is consistent with the resolved vorticity of this cell.
	
	Determine the $t_f$ for the surviving particles from the last step (if existing)
	\begin{gather}
		t_f = \text{min}\left[-\tau_n\text{ln}\left(\eta\right), \Delta t\right].
	\end{gather}
	
	\item {\bf Migration}: 
	move particles by operator splitting, which indicates
	\begin{gather}
		\vec{x}^* = \vec{x}^n + \vec{u}^n t_f,
	\end{gather}
	for free streaming, and
	\begin{gather}
		\vec{u}^{n+1} = \vec{u}^n + \vec{a} t_f, ~~~
		\vec{x}^{n+1} = \vec{x}^* + \frac{1}{2}\vec{a} t_f^2,
	\end{gather}
	for the acceleration, and meanwhile count the flux caused by the movement of particles, denoted as $\vec{w}_{i}^{fr,part}$
	\begin{gather}
		\vec{w}_{i}^{fr,part} = \sum_{k\in P\left(\partial \Omega_{i}^{+}\right)} \vec{\phi}_k - \sum_{k\in P\left(\partial \Omega_{i}^{-}\right)} \vec{\phi}_k,
	\end{gather}
	where $P\left(\partial \Omega_{i}^{+}\right)$ is the particle set moving into the cell $i$ during one time step, $P\left(\partial \Omega_{i}^{-}\right)$ is the particle set moving out of the cell $i$ during one time step, $k$ is the particle index in the set, and $\vec{\phi}_k=\left[m_{k}, m_{k}\vec{u}_k, \frac{1}{2}m_{k}\vec{u}^2_k + m_k\frac{K+3}{2} \frac{1}{2\lambda_k}\right]^T$ is the mass, momentum and energy carried by particle $k$.
	
	\item {\bf Interaction}: 
	after the migration step, binary collisions are carried out by randomly selecting two particles in one cell, following the standard DSMC collision, with strict conservation of mass, momentum, and energy.
	
	\item {\bf Annihilation}: 
	delete particles with $t_f < \Delta t$, and the carried conserved variables will merge into $\vec{W}$. 	
	Besides, calculate the TKE characterized by the surviving particles, $\rho E_t$.
\end{itemize}

Up to now, the details of evolution for wave and particle components have been introduced, and thus we have obtained all the terms of the flux, enabling the update of conserved quantities
\begin{gather}\label{particle phase equ_updateW_ugkp}
	\vec{W}_i^{n+1} = \vec{W}_i^n
	- \frac{\Delta t}{\Omega_i} \sum_{S_{ij}\in \partial \Omega_i}\vec{F}^{eq}_{ij}S_{ij}
	- \frac{\Delta t}{\Omega_i} \sum_{S_{ij}\in \partial \Omega_i}\vec{F}^{fr,wave}_{ij}S_{ij}
	+ \frac{\vec{w}_{i}^{fr,part}}{\Omega_{i}}.
\end{gather}
In addition, the update of $\vec{W}_i^{p}$ inside each cell can be obtained by summing the contributions from all particles survived inside the cell, and further $\vec{W}^h$ can be obtained based on the conservation $\vec{W}^{h}_i = \vec{W}_i^{n+1} - \vec{W}^p_i$.

In this paper, the equilibrium wall model is adopted for the turbulent boundary layer \cite{Tur-wallmodel-ode-wang2002dynamic, Tur-wallmodel-reviewARFM-bose2018wall}. The one-dimensional ordinary differential equation (ODE) for the streamwise velocity $U$ is as below
\begin{equation}\label{wm-ode}
	\frac{d}{dy} \left(\left(\mu + \mu_t\right)\frac{d U}{dy}\right) = 0,
\end{equation}
where $\mu$ is the molecular viscosity. Besides, $\mu_t$ is the eddy viscosity in wall model, and it is modeled as 
$\mu_t = \kappa \sqrt{\rho |\tau_w|} \, y \, D(y)$, with $\tau_w = \rho u_{\tau}^2$, $D(y) = \left(1 - \exp\left(-y^+/A^+\right)\right)^2
$, and model constants $\kappa = 0.41$, $A^+ = 19.0$. The above wall model including the employed $\mu_t$ model, has been extensively adopted in the study of wall-bounded turbulent flows \cite{Tur-wallmodel-transitionsensor-mettu2022wall}. 
Further, Eq.\eqref{wm-ode} is solved on a one-dimensional stretched grid spanning from the wall ($y=0$) to the chosen off-wall cell center ($y = y_c$), and the number of grid point is taken as 32 in the present study. 
In terms of coupling with WPTS simulation, the numerical flux at the wall boundary ($y=0$) is no longer computed from WPTS; instead, it is directly evaluated using the near-wall flow variables and the wall shear stress $\tau_w$ obtained from the wall-model solution.

\section{Plate transition problems}	

\subsection{Case setup and numerical validation}

As described above, in the no-particle limit, WPTS exactly reduces to the high-order GKS.
For simulations in this paper, the CFL number is taken as 0.3, and for WPTS the time step is obtained by
\begin{equation}\label{dt}
	\Delta t = \text{CFL} \times \text{Min} \left[ \frac{\Delta x}{(| \vec{U} + \delta \vec{u}| + c)}, \frac{\rho_{ref} \Delta x^2}{3 \left(\mu + \tau_t/p\right)} \right],
\end{equation}
where $\delta \vec{u}$ is the amplitude for sampling $\delta \vec{u}_p$ (particle velocity deviation from the mean value) in Eq.\eqref{velpart}, and $c$ is sound speed.

The case is a flat-plate forced transition problem with $Re=5.0\times10^4$ (using one inch as the length unit) and $Ma=0.70$, which has been well investigated by DNS \cite{Tur-case-plate-zhou2007coherent}. 
To accelerate the transition, the periodic-blowing suction disturbance in \cite{Tur-case-plate-hui2005direct, Tur-case-plate-pirozzoli2004direct} is imposed in the region $x\in[0.5,1.0]$, $y\in[0,0.1]$. At the inlet, the density and temperature are fixed as $\rho=\rho_{ref}$ and $T=T_{ref}$, with $W=0$, while the profiles of $U(y)$ and $V(y)$ are taken from the similarity solution of laminar boundary-layer at $x=30$. Besides the iso-thermal no-slip wall condition with $T_{wall}=1.098T_{ref}$ is imposed for the wall boundary, whereas the upper boundary and outlet are treated with non-reflection condition. More details about this problem can be found in the previous study \cite{Tur-case-plate-zhou2007coherent}. In this paper, two meshes are used: a fine mesh M1 ($2.75\times10^6$ cells) and a coarse mesh M2 ($2.26\times10^5$ cells), with details summarized in Table \ref{Mesh}, where the mesh R1 used in DNS is also given for comparison \cite{Tur-case-plate-zhou2007coherent}.

\begin{table}
	\begin{center}		
		\centering
		\begin{tabular}{ccccc}
			Mesh & Domain & Cell numbers & $\Delta x \times \Delta y_{w} \times \Delta z$ & $\Delta x^{+} \times \Delta y_{w}^{+} \times \Delta z^{+}$ \\
			R1  & $10.00 \times 0.65 \times 1.57$  & $1000 \times 100 \times 320$ & $0.010 \times 0.0005 \times 0.0049$ & $20.25 \times 1.013 \times 10.13$ \\
			M1  & $10.00 \times 0.50 \times 0.78$  & $520 \times 68 \times 78$ & $0.019 \times 0.0050 \times 0.010$ & $38.5 \times 10.1 \times 20.7$ \\
			M2  & $10.00 \times 0.50 \times 0.78$  & $208 \times 34 \times 32$ & $0.048 \times 0.0125 \times 0.024$ & $97.2 \times 25.3 \times 49.6$ \\
		\end{tabular}
		\caption{\label{Mesh}The details of mesh information. R1 is the mesh employed by DNS \cite{Tur-case-plate-zhou2007coherent}; M1 and M2 are the meshes employed by WM-GKS and WM-WPTS in this paper.}
	\end{center}
\end{table}

\begin{figure}
	\centering
	\begin{minipage}[t]{0.48\linewidth}
		\centering
		\includegraphics[height=5.0cm]{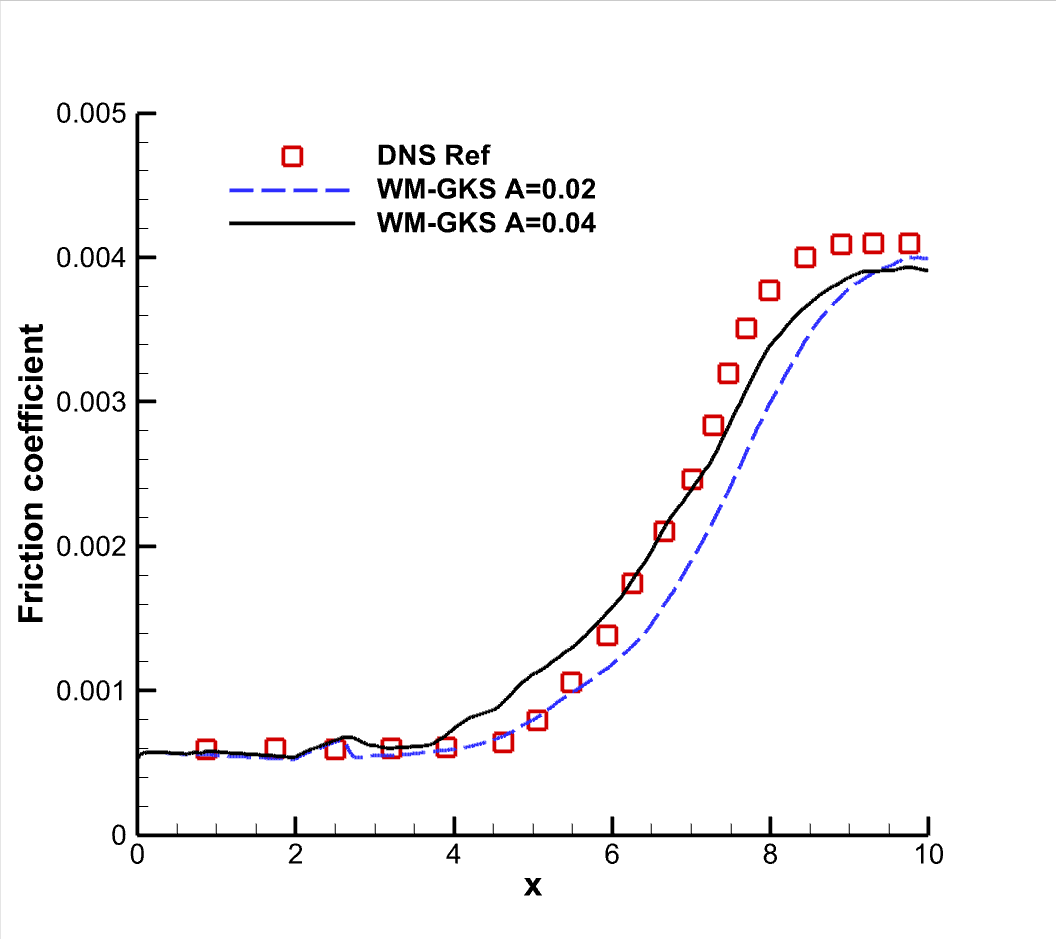}
	\end{minipage} 	
	\begin{minipage}[t]{0.48\linewidth}
		\centering
		\includegraphics[height=5.0cm]{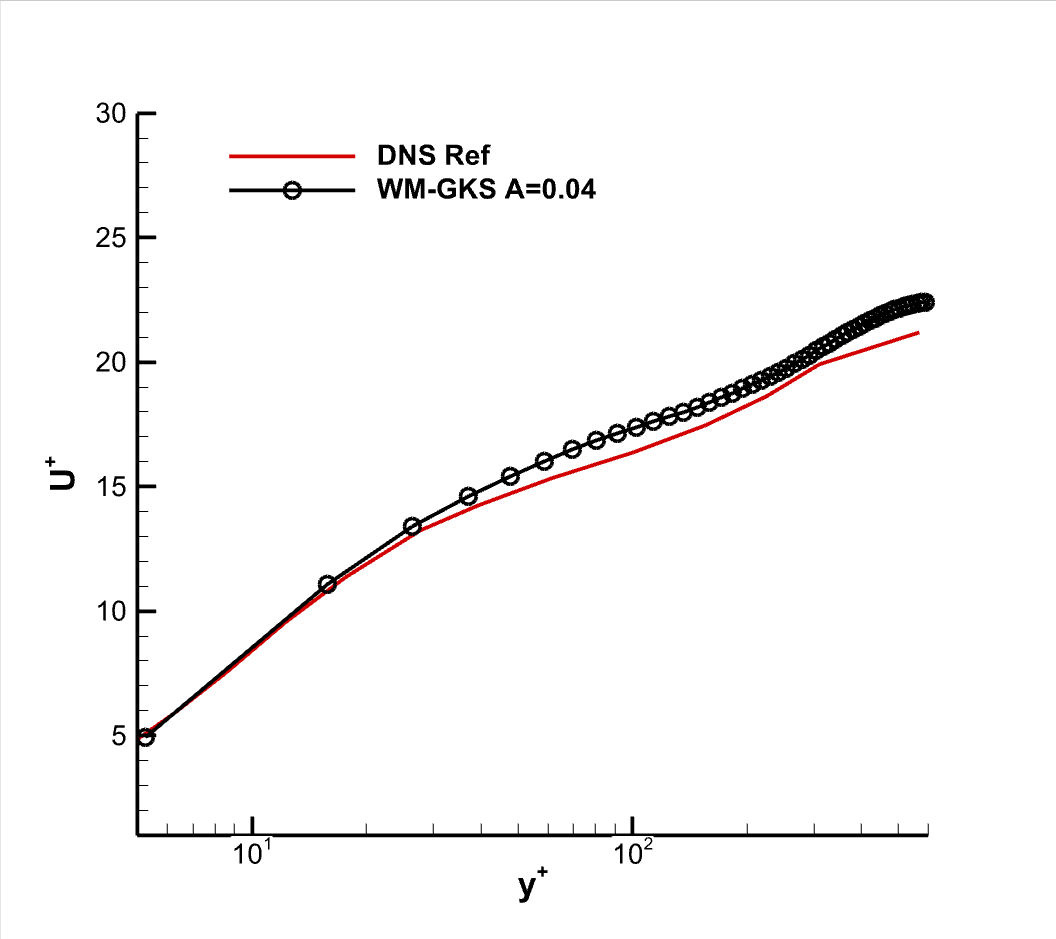}
	\end{minipage}
	\caption{(a) The friction coefficient predicted by WM-GKS under M1 grid with different disturbance amplitude $A$. (b) The mean steamwise velocity by WM-GKS with $A=0.04$.}
	\label{Fig-cf-uplu-M1}
\end{figure}

With the fine grid M1, we perform the validation by WM-GKS. The wall model is activated based on a $y^+$ threshold, with $y^+_{act}=23.4$ and the input layer taken as the 5th off-wall cell. In this paper, a weighting factor $\omega_{wm}$ is introduced to deactivate the wall model in the laminar region ($x<2$), with a linear increasing to $\omega_{wm}=1$ over $2\le x\le3$. It is worth noting that this laminar-region deactivation has no impact on the wall-model performance in the subsequent transition and turbulent zone. 
With the temporal disturbance frequency fixed at $\beta=2.5$, different disturbance amplitudes $A$ are tested. As shown in Fig.\ref{Fig-cf-uplu-M1}(a), the friction coefficient $c_f$ obtained with $A=0.04$ agrees better with the DNS reference, capturing both the transition process satisfactorily. Further, the mean streamwise velocity profile at $x=9.9$, presented in Fig.\ref{Fig-cf-uplu-M1}(b), also exhibits excellent agreement with the DNS data. These results validate that, by the fine grid M1, performs well the GKS including the wall model and employed disturbance with $A=0.04$, which will be adopted for the subsequent comparisons on coarse grid.

\subsection{Performance of WM-GKS and WM-WPTS on coarse grid}

\begin{figure}
	\centering
	\begin{minipage}[t]{0.48\linewidth}
		\centering
		\includegraphics[height=5.0cm]{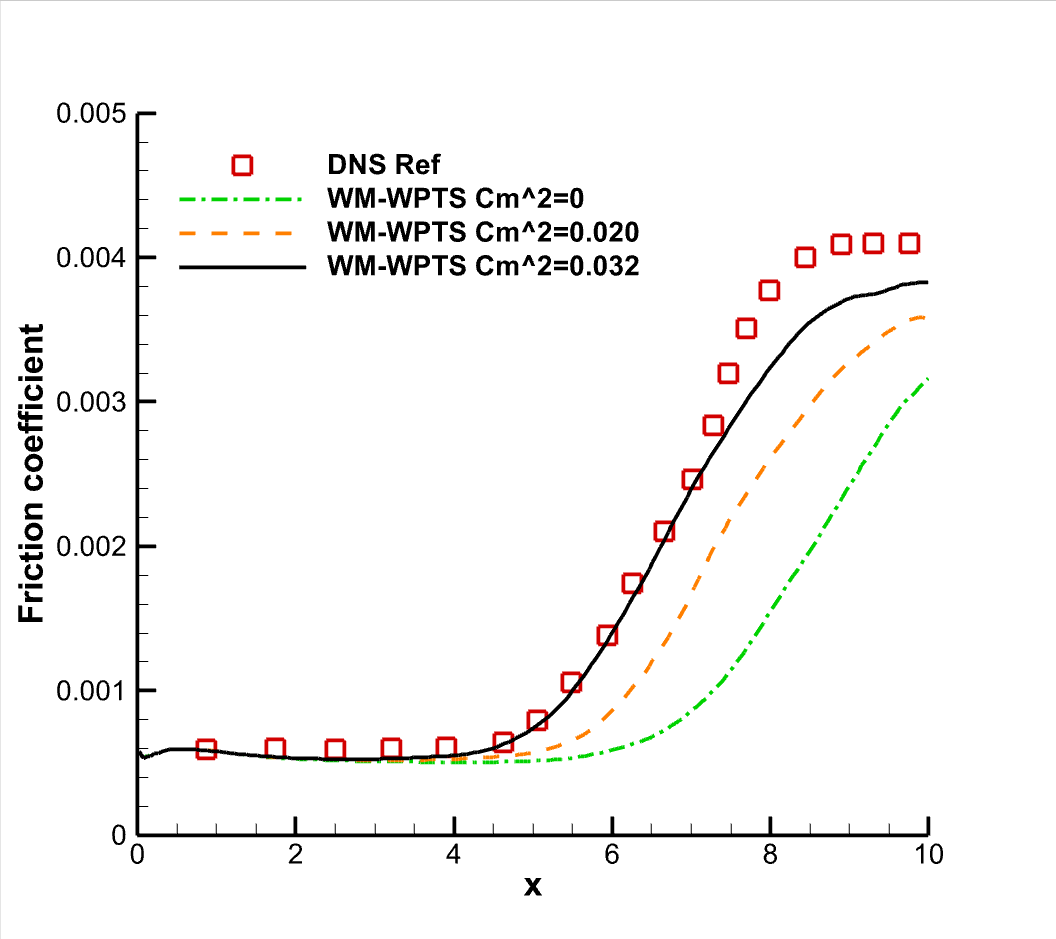}
	\end{minipage} 	
	\begin{minipage}[t]{0.48\linewidth}
		\centering
		\includegraphics[height=5.0cm]{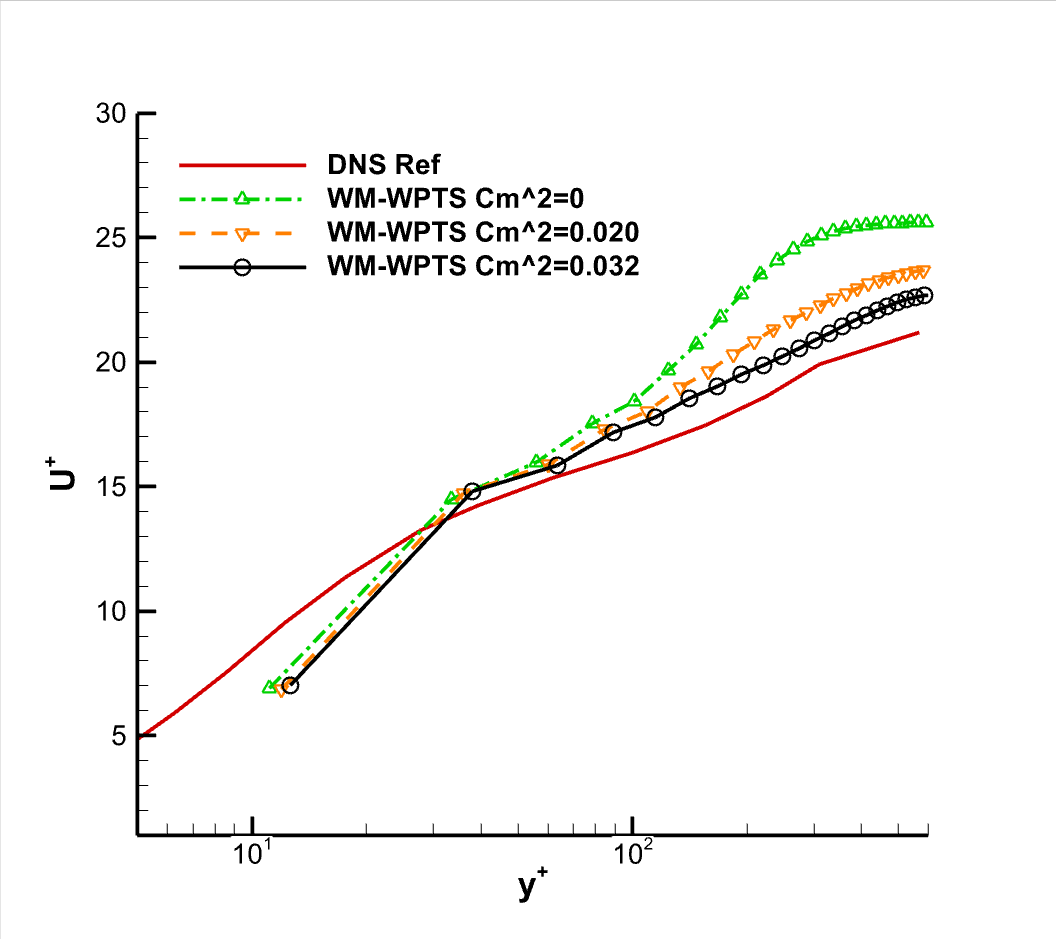}
	\end{minipage}
	\caption{(a) The friction coefficient predicted by WM-WPTS under coarse grid M2 with different $C_m^2$. (b) The corresponding mean steamwise velocity by WM-WPTS. Note that WM-WPTS will reduce to WM-GKS when $C_m^2=0$.}
	\label{Fig-cf-uplu-M2}
\end{figure}

As shown in Table \ref{Mesh}, the minimum near-wall grid spacing in the coarse grid M2 is $2.5$ times larger than that of M1. To maintain the consistency of the wall model activation in plate transition problems, the threshold is equivalently converted to $y^+_{act}=19.5$, with the input layer shifted to the 2nd off-wall cell.
Firstly the performance of WM-GKS is evaluated on M2, and the results are presented in Fig.\ref{Fig-cf-uplu-M2} labeled with ``WM-WPTS $C_m^2$=0'' (green color), since when there is no particles with $C_m^2=0$ WM-WPTS will reduce to WP-GKS. 
Fig.\ref{Fig-cf-uplu-M2}(a) shows the $c_f$ curve shifts noticeably to the right relative to the reference, indicating a significant delay in transition onset. Besides, the mean velocity profile at $x=9.9$ given in Fig.\ref{Fig-cf-uplu-M2}(b) also exhibits obvious discrepancies, suggesting that the flow has not yet reached a fully turbulent state at this location. 
The above results indicate WM-GKS is incapable of accurately predicting the flow transition under such under-resolved condition.

Then WM-GKS is replaced by WM-WPTS, while keeping the same all other settings. WPTS introduces discrete particles with non-equilibrium transport in the turbulent zone, where the proportion of particle component and their transport time are overall governed by $\tau_n = \tau + \tau_t$. In this paper, $\tau_t = \rho \nu_t / p$ and the $\tau_t$ is obtained by the mixing-length-based model
\begin{equation}\label{nut}
	\nu_t = \alpha\left(\rho\right) C_m^2 y^2 |\vec{S}| D_{srs}\left(y\right),
\end{equation}
where $y$ is distance to the wall, and $|\vec{S}|$ is the magnitude of strain rate. Besides, $D_{srs}\left(y\right)=\left[\text{exp}\left(-y/y_{ref}\right)\right]^2$ with $y_{ref}=0.15$, and the term $D_{srs}(y)$ originates from the distinct treatment of the mixing length in scale-resolving simulations, such as WPTS, versus RANS approaches where all turbulent scales are modeled. $\alpha\left(\rho\right)$ is the high-density modification, taken as $\alpha \left(\rho \right) = 1.0 /\text{max} \left[1.0, 1.0+\text{1.0E2}\left(\rho - \rho_{ref}\right)\right]$.
Fig.\ref{Fig-cf-uplu-M2} presents the $c_f$ distributions predicted by WM-WPTS with different $C_m^2$ values. At $C_m^2=0.032$, the transition delay is substantially alleviated, and the $c_f$ curve agrees considerably with the DNS data in terms of onset location, and the transitional shape. Although perfect alignment is not achieved, WPTS markedly improves the prediction of key transition characteristics compared to WM-GKS under such a coarse grid. Furthermore, at $C_m^2=0.032$ the mean velocity profile shown in Fig.\ref{Fig-cf-uplu-M2}(b) recovers a well-defined logarithmic layer, confirming the capability of WPTS to restore turbulent statistical properties.
Building upon this validation, the following shows some evolutionary characteristics of the flow field, along with the feature of wave-particle numerical method.

\begin{figure}
	\centering
	\begin{minipage}[t]{0.48\linewidth}
		\centering
		\includegraphics[height=3.0cm]{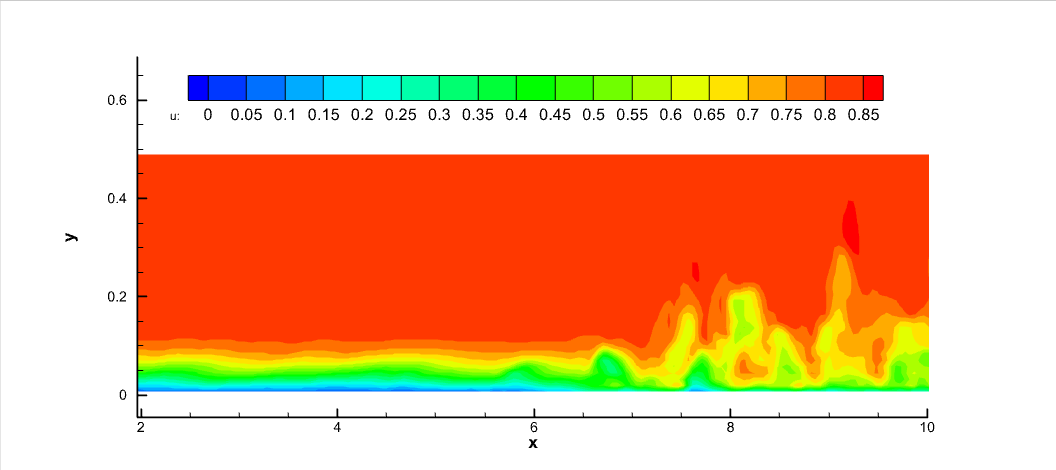}
	\end{minipage} 
	\begin{minipage}[t]{0.48\linewidth}
		\centering
		\includegraphics[height=3.0cm]{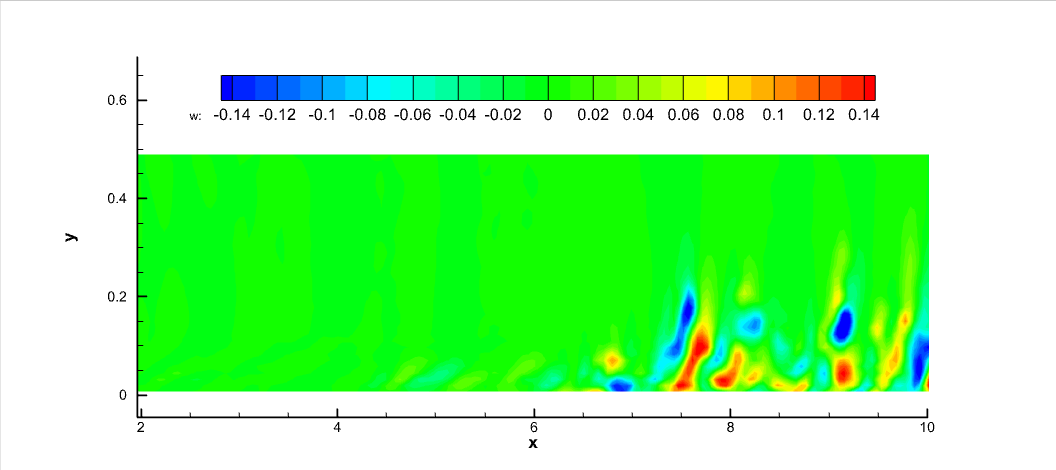}
	\end{minipage}	
	\begin{minipage}[t]{0.48\linewidth}
		\centering
		\includegraphics[height=3.0cm]{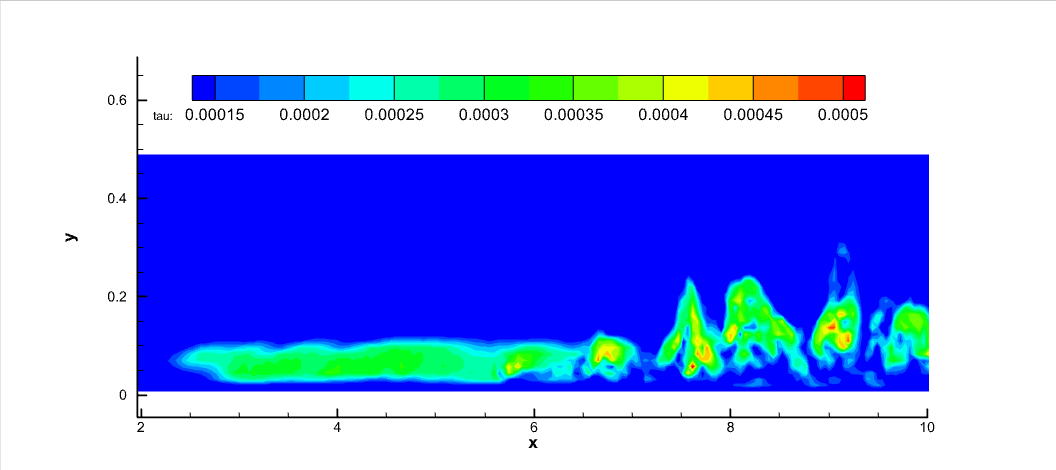}
	\end{minipage}
	\begin{minipage}[t]{0.48\linewidth}
		\centering
		\includegraphics[height=3.0cm]{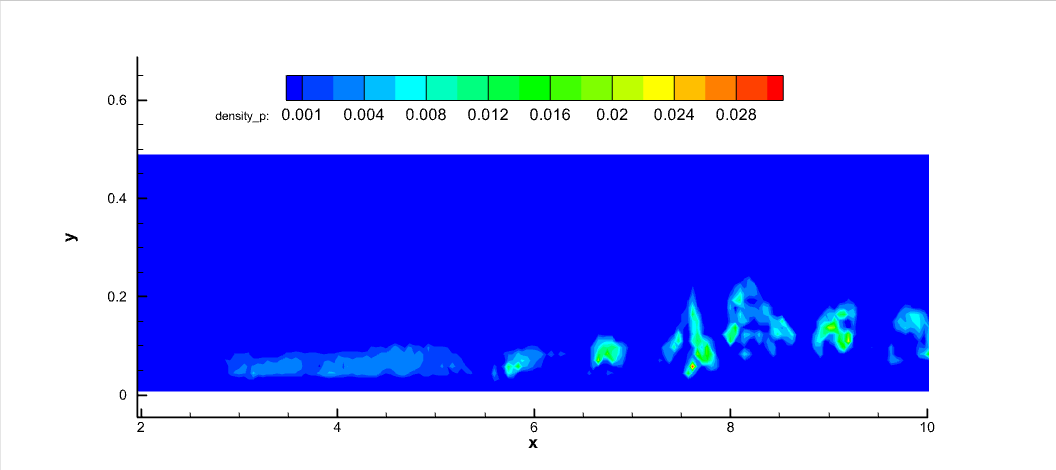}
	\end{minipage}
	\caption{The predicted snapshots by WM-WPTS under coarse grid M2: (a) streamwise velocity $U$, (b) spanwise velocity $W$, (c) characteristic time $\tau_n$, and (d) the fluid density of surviving particles on the $z=0.39$ plane. The domain shown for $x$ is $[2.0,10.0]$.}
	\label{Fig-ins-M2}
\end{figure}

Specifically, Fig.\ref{Fig-ins-M2} presents instantaneous snapshots on the $z=z_{mid}$ plane predicted by WM-WPTS, including the streamwise velocity $U$, the spanwise velocity $W$, the characteristic time $\tau_n$, and the fluid density by surviving particles. 
Prior to the onset of transition, a limited number of stochastic particles emerge due to shear-induced mechanisms. As the flow undergoes transition and subsequently becomes fully turbulent during its downstream evolution, Fig.\ref{Fig-ins-M2}(c) exhibits a clear increase in the modeled characteristic time scale $\tau_n$, alongside a growing proportion of the particle component within the WPTS shown in Fig.\ref{Fig-ins-M2}(d), indicating intensified non-equilibrium transport of discrete fluid elements. Although not captured in the instantaneous flow snapshots, these stochastic particles dynamically annihilate and regenerate in response to the local macroscopic flow structures. This dynamic behavior compensates for the loss of small-scale flow information that would otherwise be under-resolved due to insufficient grid resolution, thereby enhancing the predictive accuracy for transitional flows.

\section{Conclusion}

In this paper, the model equations of WPTS have been explicitly formulated and the underlying physical mechanisms of each term are interpreted, establishing a solid theoretical foundation for the method. Further, the WPTS is coupled with a wall model (WM-WPTS) for wall-bounded turbulent flows and applied to the flat-plate transition problem, a canonical case featuring multiple flow regimes including laminar, transitional, and fully turbulent states. 
Under a coarse grid with approximately 0.2 million cells, the conventional WM-GKS (a kinetic NS solver) exhibits a significant delay in predicting transition onset. In contrast, with the choice appropriate coefficient for $\tau_t$, the WM-WPTS yields both the wall friction coefficient and the mean streamwise velocity profiles in the turbulent zone that closely match DNS data by grid with around 10 million points. This improvement can be attributed to the introduction of stochastic particles with non-equilibrium transport, which enhances the predictive capability on coarse meshes. 
These results indicate that the WM-WPTS represents a novel and promising tool for wall-bounded turbulent flow simulations.

\section{Acknowledgements}
The current research is supported by National Key R\&D Program of China (Grant Nos. 2022YFA1004500), National Science Foundation of China (92371107), and Hong Kong research grant council (16208324).

\bibliographystyle{plain}%
\bibliography{jfm}
\end{document}